\documentclass{PoS}

\title{One dimensional supersymmetric Yang-Mills theory with 16 supercharges}

\ShortTitle{One dimensional supersymmetric Yang- Mills theory with 16 supercharges}

\author{\speaker{Daisuke Kadoh}\\%
        Quantum Hadron Physics Laboratory, RIKEN,
2-1 Hirosawa, Wako, Saitama 351-0198, Japan\\
        E-mail: \email{kadoh@riken.jp}\\
}

\author{Syo Kamata\\
Graduate School of Science, Rikkyo University, 3-34-1 Nishi-Ikebukuro, Toshima-ku,
Tokyo 171-8501, Japan\\
 E-mail: \email{skamata@rikkyo.ac.jp}
}

\abstract{
We report on numerical simulations of one dimensional maximally supersymmetric SU(N) Yang-Mills theory,
 by using the lattice action with two exact supercharges.  
Based on the gauge/gravity duality, the gauge theory corresponds to N D0-branes system 
in type IIA superstring theory at finite temperature.
We aim to verify the gauge/gravity duality numerically by comparing our results of the gauge side with analytic solutions of the gravity side.
First of all, by examining the supersymmetric Ward-Takahashi relation, we show that supersymmetry breaking effects 
from the cut-off vanish in the continuum limit and our lattice theory has the desired continuum limit.
Then, we find that, at low temperature, the black hole internal energy obtained from our data is close to the analytic solution of the gravity side. It suggests the validity of the duality.
}

\FullConference{The 30 International Symposium on Lattice Field Theory\\
Cairns, Australia\\
June 24-29, 2012}

\begin{document}

\section{Introduction}

Gauge/gravity duality asserts an equivalence 
between strongly coupled gauge theory and the classical gravity on curved space, 
which was originally stated as AdS/CFT correspondence
which includes the supersymmetry by Maldacena\cite{Maldacena:1997re}.
From the duality, we expect that strongly coupled gauge theories, 
which are usually difficult to calculate by hand, can be analytically solved
via the gravity side.
So, there are many applications from the context, getting over the barrier among fields (for example, 
elementary particle physics, cosmology, condensed matter physics and so on).
However, it is a conjecture and therefore verifying the duality in some way is desirable.

We aim to verify the gauge/gravity duality from lattice simulations 
in one dimensional supersymmetric Yang-Mills theory with sixteen supercharges. 
The theory is obtained by dimensional reduction from 10d ${\cal{N}}=1$ SYM (or 4d ${\cal{N}}=4$ SYM). 
Actually, we introduce temperature into the theory 
by imposing the anti-periodic boundary conditions on fermions. 
Based on the duality, the gauge theory corresponds to N D0-branes system 
in type IIA superstring theory at finite temperature.
In particular, at low temperature, the gauge theory becomes strong coupling and 
using analytic techniques to examine the dual black hole physics from gauge theory become difficult.
So, we use the lattice gauge theory to analize the gauge theory.
From comparisons between lattice results and analytic solutions of the gravity side, 
we discuss the validity of the gauge/gravity duality.

There are two previous works about numerical simulations of the 1d maximally supersymmetric 
Yang-Mills theory: non-lattice simulations done 
by Nishimura et al.\cite{Anagnostopoulos:2007fw,Hanada:2008gy, Hanada:2008ez,Hanada:2009ne}
and lattice simulations done by Catterall and Wiseman\cite{Catterall:2007fp}.
%,Catterall:2008yz,Catterall:2009xn}.
Supersymmetry is broken by the cut-off effects in their regularized theories. 
Nevertheless, both results are consistent with the gravity side from UV-finiteness of 1d gauge theories. 
In contrast, we employ the lattice formulations with a few supersymmetric charges on the lattice
\cite{Sugino:2004gz,Kaplan:2005ta},
which have been recently developing, 
in particular, our lattice theory has two exact supercharges even on the lattice.
We expect that the lattice theory has some advantages, for example, clear signals etc., thanks to the exact charges, 
in high accuracy verifications of the duality.

In section 2 we explain our lattice theory 
and then in section 3 we see some details of simulation techniques. 
In section 4 we show that the lattice theory 
has the correct continuum limit by computing the supersymmetric Ward-Takahashi relation.
In section 5  we show the internal energy of the dual black hole obtained from gauge theory side 
as an evidence of the gauge/gravity duality.

\section{1d SYM with 16 supercharges}

The supersymmetric $SU(N)$ Yang-Mills theory with sixteen supercharges 
is a gauge theory in which a gauge field of the temporal direction $A_1$ 
interacts with nine scalar fields $X_i(i=1,\cdots,9)$ and real sixteen fermions
$\psi_\alpha \ (\alpha=1,\cdots,16)$.
The continuum action is given by 
\begin{eqnarray}
%S_{\mathrm{SYM}}^{1D} 
S_{cont}
&=& \frac{N}{\lambda} \int dt \;  \mathrm{tr}\left\{
\frac{1}{2} 
(D_{1} X_{i}(t))^{2} - 
%\left( 
\frac{1}{4}   [X_{i}(t),X_{j}(t)]^2 \right. \nonumber \\
&& \hspace{2cm} \left. %\right)^{2} 
+  \frac{1}{2} \psi^T(t) D_{1} \psi(t) 
+ \frac{1}{2}  \psi^T(t)  \gamma_{i} 
[X_{i}(t), \psi (t)]  \right\},  
 \label{eq.SYM_action} 
\end{eqnarray}
where $\lambda$ is the 't Hooft coupling constant. 
Here, all fields are expanded as
$\varphi(t)=\sum_{a=1}^{N^2-1}\varphi^a(t) T^a$ by gauge group generators $T^a$ of the $SU(N)$ group
\footnote{
The generators $T^a$ satisfy the normalization condition, $\mathrm{tr}(T^{a}T^{b})=\delta_{ab}$.}
.
Also, the covariant differential operator $D_1$ is
defined through $D_1 \varphi=\partial_1 \varphi+i[A_1,\varphi]$.
The $\gamma_i$ are real symmetric matrices which satisfy the nine dimensional Euclidean Clifford algebra.

The realization of supersymmetry on the lattice has been a difficult issue
due to the lack of Leibniz rule on the lattice for a long time. 
However, recently, 
Sugino proposed a lattice formulation of maximally supersymmetric Yang-Mills theory 
with two exact supercharges\cite{Sugino:2004gz}
from the topological twisted version \cite{Vafa:1994tf}.
In the twisted theory,  the original action eq.(\ref{eq.SYM_action}) can be rewritten 
as a closed form, $S_{cont}=Q_+Q_-(...)$, using two supercharges $Q_{\pm}$
where $Q_{\pm}^2$ are gauge transformations. 
From the nilpotency of $Q_{\pm}$ up to gauge transformations, we can see that 
the action is invariant under $Q_{\pm}$-transformations, 
without the obvious use of Leibniz rule.

Let us consider one dimensional lattice of the size $L$ with the periodic boundary condition.
Scalars and fermions are defined on sites labeled by $t=0,\cdots,L-1$ while a gauge field is defined on links 
through the link field $U_1$ to realize the exact gauge invariance. 
Our lattice action is defined by
\begin{eqnarray}
S &=& Q_{+} Q_{-} \frac{N}{2 \lambda_{0}} \sum_{t=0}^{L-1} \, {\rm tr} 
\left[ -i \sum_{i=1}^3 B_{i}(t) \Phi_{i}(t) -\frac{1}{3} \sum_{i,j,k=1}^3 \epsilon_{ijk} B_{i}(t) [B_{j}(t),B_{k}(t)] 
\right. \nonumber \\
 && \hspace{3cm}
\left. - \sum_{\mu=1}^{4}\psi_{+\mu}(t) \psi_{-\mu}(t) 
- \sum_{i=1}^3\xi_{+i}(t) \xi_{-i}(t) -\frac{1}{4} \eta_{+}(t) \eta_{-}(t) \right],
%\qquad i,j,k=1,2,3, \mu=1 ,\cdots ,4 .
\label{eq:lat_action}
\end{eqnarray}
where $\lambda_0$ is a dimensionless 't Hooft coupling constant defined by $\lambda_0=\lambda a^3$
with the lattice spacing $a$.
Here, $B_i,C,\phi_\pm$ and $\psi_{\pm\mu},\eta_{\pm},\xi_{\pm i}$ are
some combinations of original scalars and those of fermions, respectively 
(our notation follows \cite{Sugino:2004gz}, or see \cite{Kadoh:2012a}).
Lattice $Q_{\pm}$-transformations are defined over the new variables
\footnote{The transformations here are those for $U_1$ and $\psi_{\pm1}$ (for the others, see \cite{Kadoh:2012a} ).},
\begin{eqnarray}
 Q_{\pm}U_{1}(t) = i \psi_{\pm 1}(t)U_{1}(t), \qquad
 Q_{\pm} \psi_{\pm 1}(t) = i\psi_{\pm 1}(t) \psi_{\pm 1}(t) \pm i \nabla_{1} \phi_{\pm}(t) ,\quad
  \cdots \ \ . 
\end{eqnarray}
From the definitions above, $Q_{\pm}^2$ are
gauge transformations with gauge parameters $\phi_\pm,C$, 
\begin{eqnarray}
Q_{+}^{2}=\delta_{-i \phi_+}, \quad Q_{-}^{2}=\delta_{i \phi_-}, \quad \{ Q_{+},Q_{-} \} = \delta_{-i C},
\label{2q}
\end{eqnarray}
where $\delta_\omega$ is a gauge transformation with the parameter $\omega$.
As a result, $Q_{\pm}$-invariance is realized even on the lattice, because of eq.(\ref{eq:lat_action}) 
and the exact gauge invariance of the lattice theory.

The continuum limit is realized by taking $\lambda_0$ to zero while keeping a typical scale of this system
(e.g., the dimensionful 't Hooft coupling $\lambda$). Also, the lattice action has no doublers because $(D^\dag_0 D^{}_0)(p)=4 {\rm sin}^2(p_1/2)$
where $D^{}_0$ is the free limit of the lattice Dirac operator.

To introduce temperature,
we change the boundary condition on fermions from periodic one to anti-periodic one,
while keeping that on bosons,
\begin{eqnarray}
&& U_1(t)=U_1(t+L),\qquad 
X_{i}(t)=X_{i}(t+L), \quad (i=1,\cdots,9),\\ \qquad
&& \psi_\alpha(t)=-\psi_\alpha(t+L), \quad (\alpha=1,\cdots,16),
\end{eqnarray}
where temperature is defined by $T=1/(La)$.
Temperature breaks all supersymmetries explicitly and, of course, $Q_{\pm}$-invariance.
Hereafter, $T$ denotes a dimensionless temperature $T/\lambda^{1/3}$.

\section{Simulation details}

We use the standard Hybrid Monte Carlo method. But, 
there are two additional difficulties in our fermion sector: 
the 4-fermi interaction and the pfaffian, as explained below.

After $Q_{\pm}$-transformations, the action eq.(\ref{eq:lat_action}) includes a cut-off order
4-fermi interaction as
\begin{eqnarray}
Q_+ Q_- \mathrm{tr} \left(\psi_{+1} \psi_{-1} \right) 
\sim 
\mathrm{tr} \left( \left\{\psi_{+1}, \psi_{-1} \right\}^2 \right)
\quad \longrightarrow \quad \mathrm{tr} \left(\sigma^2 + \sigma \left\{\psi_{+1}, \psi_{-1} \right\} \right).
\label{eq:sigma}
\end{eqnarray}
For the 4-fermi interaction, we introduce an auxiliary field $\sigma$ to write it as the third term
in eq.(\ref{eq:sigma}). We treat ${\rm tr} (\sigma^2)$ as a part of the boson action and 
${\rm tr}(\sigma \{\psi_{+1},\psi_{-1}\})$ as a part of the fermion bilinear $S_F=\sum_t \psi^T(t) D \psi(t)$,
without integrating $\sigma$. So, we regard 10+1 bosonic fields, $U_1,A_2,\cdots,A_{10},\sigma$, 
as configurations generated by HMC method.

The integration of fermions becomes the pfaffian, ${\rm pf}(D)$, which generally takes complex values.  
We treat the absolute value and the complex phase of the pfaffian, individually, to avoid the sign problem. 
The absolute value can be given as an integral by pseudo fermion $\phi$ since $|{\rm pf}(D)|={\rm det}(D^\dag D)^{1/4}$, and the 4th root is approximately by the rational expansion,   
\begin{eqnarray}
\displaystyle
|{\rm pf}(D)| = \int D\phi^\dag D \phi\, {\rm exp}\left\{
- \sum_{t=0}^{L-1} \phi^\dag(t) \left[\alpha_0 + \sum_{i=1}^M \frac{\alpha_i}{D^\dag D + \beta_i}  \right]\phi(t)
\right\},
\label{eq:rational_approx}
\end{eqnarray}
where the order $M$ and the coefficients $\alpha_i,\beta_i$ of the approximation are 
determined from the range of $D^\dag D $'s eigenvalues measured in the simulations.
We compute inversions of $D^\dag D$  with shifts $\beta_i$ in eq.(\ref{eq:rational_approx}) 
by using multi-mass shifted solver. Also, for the phase of pfaffian, 
we use the phase quench, or use the phase reweighting method if we want to include the effect in the results.

For $T \gg 1$, HMC method stably works and we can obtain sufficient statistics. 
However, as temperature decreases, at some temperature (which actually depend on $N$), 
the magnitudes of scalar fields monotonically increase
against Monte Carlo trajectories and therefore the thermalization does not occur. 
This instability is related to the classical flat direction of the boson action. 
To avoid the instability, in parameter regions where run-away modes to the flat direction appear,
we introduce a mass term of scalar fields,  
\begin{eqnarray}
S_{\mathrm{mass}} &=& \mu^{2}_{0}\, \frac{N}{2\lambda_0}\,  \sum_{t=0}^{L-1} \; \sum_{i=1}^{9} 
 \mathrm{tr} \left( X^{2}_{i} (t) \right), \label{eq:mass}
%S_{\mathrm{mass}} &=& \frac{N \mu^{2} }o2\lambda} \int dt \; \sum_{\nu=2}^{10}\mathrm{tr} \left[ A^{2}_{\nu} \right], \nl
\end{eqnarray}
where $\mu_0$ is a dimensionless mass. Hereafter $\mu$ denotes a dimensionless mass $\mu/\lambda^{1/3}$.

\section{Supersymmetric Ward-Takahashi relation}

\begin{figure}[t]
\begin{center}
\hspace{-0.6cm}
\includegraphics[width=7cm,keepaspectratio,clip]{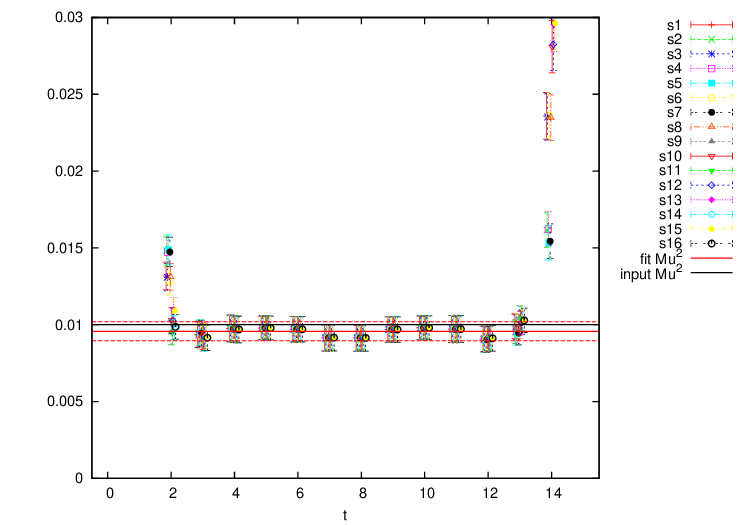}
\hspace{0.6cm}
\includegraphics[width=7cm,keepaspectratio,clip]{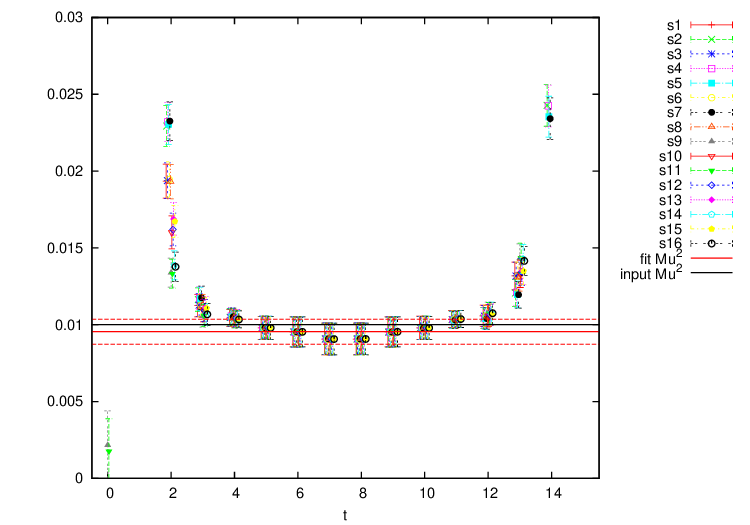}
\caption{The ratio $\frac{<\partial_1 J_1(t)Y(0)>}{<Y(t)Y(0)>}$ is plotted against the temporal direction 
for $N=3$ (Left) and $N=4$ (Right) on the $L=16$ lattice. 
The labels $s1$, $\cdots$, $s16$ denote the spin indices of supercurrent $J_1$.
Red lines represent the fitted masses obtained by fitting plateaus.
They are consistent with the mass, $\mu^2$ (black lines) within statistical errors. 
Deviations from the plateaus near the lattice boundary correspond to the contact terms.
}
\label{fig:swti}
\end{center}
\end{figure}

Supersymmetry is broken by the lattice cut-off. 
In the classical continuum limit, the breaking effects by the cut-off identically vanish.
By contrast, in the quantum theory, it is not so clear whether the breaking effects generally vanish
in the continuum limit because of ultra violet divergences and non-perturbative effects.
Fortunately, the 1d gauge theory is UV-finite and any operators which break supersymmetry
are not generated radiatively, so our lattice theory has the correct continuum limit, at least, in the perturbation theory.  However, we do not know a-priori
whether the similar argument is possible beyond the perturbation theory.
Also, we must divide the cut-off effects from the other breaking sources, 
the temperature and the mass term, eq.(\ref{eq:mass}).

For the issue, Kanamori and Suzuki used a method
which can extract only the cut-off effects in lattice simulations of 2d ${\cal N}=(2,2)$ SU(2) SYM\cite{Kanamori:2008bk}. 
The method is a simple one measuring of the partially conserved SUSY current on the lattice.
We use the same method and check whether the cut-off effects vanish in the continuum limit 
without relying on the perturbation theory.

For the continuum theory with mass term,
we have the partially conserved supersymmetric Ward-Takahashi relation, 
\begin{eqnarray}
\left\langle \partial_{1} J_{1 \alpha} (t) \mathcal{O}(s) \right\rangle 
&=& \mu^{2} \left\langle  Y_{\alpha} (t)  \mathcal{O}(s) \right\rangle 
- \delta(t-s) \left\langle  Q_{\alpha} \mathcal{O}(s) \right\rangle,
\label{eq:pcsc}
\end{eqnarray}
where ${\cal O}$ is an arbitrary operator and $Q_\alpha (\alpha=1,\cdots,16)$ are supercharges 
which generate supersymmetric transformations. Here $J_{1}$ and $Y$ are 
the supercurrent and the breaking term from mass term, respectively, which are defined as 
\begin{eqnarray}
J_{1\alpha } &=& \frac{N}{\lambda} 
 \left[
   \sum_{i=1}^{9} \sum_{\beta=1}^{16}  (\gamma_i)_{\alpha\beta} \ \mathrm{tr} \left(\psi_\beta D_1 X_i \right)
   +\frac{1}{2} \sum_{i,j=1}^{9}\sum_{\beta=1}^{16}\ (\gamma_i \gamma_j)_{\alpha\beta} \  
     \mathrm{tr} ( \psi_{\beta} [X_i,X_j] ) 
 \right], \\
Y_\alpha &=& \frac{N}{\lambda} \sum^{9}_{i=1} \sum_{\beta=1}^{16} 
(\gamma_i)_{\alpha\beta} \
\mathrm{tr} \left\{ X_i \psi_\beta \right\}.
\end{eqnarray}
%where $\Gamma_M$ are gamma matrices in 10 dimensions.

In the continuum theory, the supersymmetric Ward-Takahashi relation, eq.(\ref{eq:pcsc}), 
holds even at finite temperature.
It means that we can find the cut-off effects by measuring a lattice counterpart of the relation. 
Actually, we compute the following ratio, 
\begin{eqnarray}
\frac{ <\nabla^S_1 J_{1\alpha}(t) Y_\beta(0)> }{<Y_\alpha(t)Y_\beta(0)>}, \qquad {\rm for}\ \ \alpha, \beta=1,\cdots,16, \label{eq:ratio}
\end{eqnarray}
where $\nabla^S_1$ is the symmetric covariant difference operator. 
Here we only use the forward covariant difference operator in the lattice definition of the supercurrent $J_1$.
Also, for fixed $\alpha$, $\beta$ is uniquely determined because correlators in the denominator with other $\beta$ are nearly zero, that is, the ratio is meaningless. If the cut-off effects vanish in the continuum limit, 
the ratio must be $\mu^2$ 
in the limit, from eq.(\ref{eq:pcsc}).

In Figure \ref{fig:swti}, we plot the ratio for $N=3$ and $N=4$ with $T=1$, $\mu^2=0.01$ 
and the lattice size $L=16$. 
The horizontal axis represents the temporal direction.  
The corresponding lattice spacing is $a=0.0625$ in the unit of $\lambda=1$.
For both cases, clear plateaus are observed within statistical errors. 
We perform the constant fit for the plateaus. The obtained values are consistent with
$\mu^2=0.01$ within the statistical errors.
The result suggests that cut-off effects mostly vanish near the continuum limit and
our lattice theory has the correct continuum limit beyond the perturbation theory 
for, at least, $T=1$ of $N=3$ and $N=4$.

\section{Internal energy}

\begin{figure}[t]
\begin{center}
\hspace{-0.6cm}
\hspace{1cm}
\includegraphics[width=8.5cm,keepaspectratio,clip]{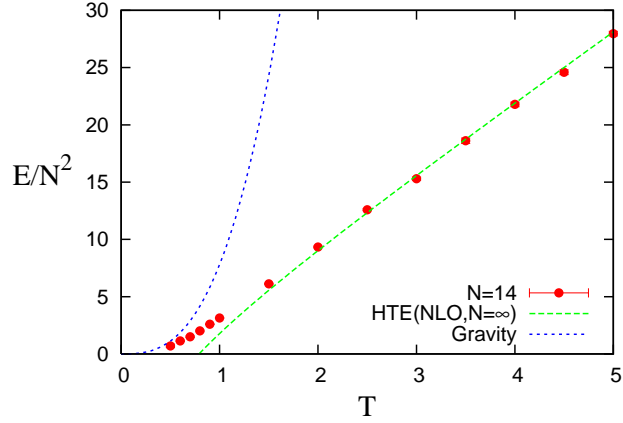}
\caption{
The internal energy of black hole, normalized by $N^2$.
Data points(red circles) are our results for $N=14$.
The dashed green line corresponds to the result obtained by high temperature expansion 
at next to leading order in the large N limit. 
The dashed blue curve represents the analytic solution of the gravity side.}
\label{fig:energy}
\end{center}
\end{figure}

The internal energy of the black hole, associated with the black hole thermodynamics, 
is one of simple examples to test the gauge/gravity duality. 
In the gravity side,
the internal energy $E$ is related to temperature $T$ through an analytic formula\cite{Klebanov:1996un},
\begin{eqnarray}
\frac{1}{N^2}\left(\frac{E}{\lambda^{1/3}}\right) = c \left(\frac{T}{\lambda^{1/3}}\right)^{14/5},
\qquad c=\frac{9}{14}\left\{4^{13} 15^2 \left(\frac{\pi}{7}\right)^{14}\right\}^{1/5} = 7.41 \cdots \ \ .
\label{energy_gravity}
\end{eqnarray}
We compute $E=-\frac{\partial}{\partial \beta} {\rm ln}Z$, where
$\beta$ is the inverse temperature, from our data and compare it with the above analytic formula.

In Figure \ref{fig:energy}, we show the internal energy versus temperature
for $N=14$ and $\mu^2=0$. We used two different lattice sizes, $L=8$ for $T\ge 1$ and $L=16$ for $T < 1$.
For high temperature, we see that our data and the result obtained by high temperature expansion at next to leading order\cite{Kawahara:2007ib} agree as expected. However, as temperature decreases, our data departs from the curve of the high temperature expansion around $T=1.5$ and it is smoothly close to the analytic curve of the gravity side,
eq.(\ref{energy_gravity}). This suggests the validity of the gauge/gravity duality in this system.

In $N=14$, 
HMC-runs are stable for $T \ge 0.5$ without the mass term, however,
for $T < 0.5$, the instability from the flat direction occurs. So, 
we must use $N >14$ to explore possible further lower temperature.
Also, simulations with different lattice spacings at same temperature are necessary
to take the continuum limit. 
Such simulations to obtain high accuracy results are in progress\cite{Kadoh:2012b}.

\end{document}